\documentclass[aps,prl,reprint,twocolumn,showpacs,superscriptaddress,amsmath,amssymb]{revtex4-1}

\usepackage{graphicx}  % needed for figures5
\usepackage{dcolumn}   % needed for some tables
\usepackage{bm}        % for math
\usepackage{hyperref}
\usepackage{verbatim}  % for comment
\usepackage{natbib}
\usepackage{epsfig}
\usepackage{epstopdf}
\usepackage{booktabs}  % for multicolumn 
\usepackage{color}
\usepackage[toc,page]{appendix}
\usepackage{placeins}
\usepackage{float}
\usepackage{multirow}
\usepackage[normalem]{ulem}
\usepackage{amsmath}
\usepackage{tabu}
\usepackage[utf8]{inputenc}
\usepackage{rotating}
\usepackage{multirow}
\usepackage[export]{adjustbox}
\begin{document}

%\widetext
\title{Ternary free-energy entropic lattice Boltzmann model with high density ratio}

\author{M. W\"{o}hrwag}
\affiliation{Department of Physics, Durham University, South Road, Durham DH1 3LE, UK}
\affiliation{Department of Mechanical and Process Engineering, ETH Zurich, CH-8092 Zurich, Switzerland}
\author{C. Semprebon}
\affiliation{Smart Materials \& Surfaces Laboratory, Northumbria University, Newcastle upon Tyne NE1 8ST, UK}
\author{A. Mazloomi M.}
\affiliation{Department of Mechanical and Process Engineering, ETH Zurich, CH-8092 Zurich, Switzerland}
\author{I. Karlin}
\email{Email: karlin@lav.mavt.ethz.ch}
\affiliation{Department of Mechanical and Process Engineering, ETH Zurich, CH-8092 Zurich, Switzerland}
\author{H. Kusumaatmaja}
\email{Email: halim.kusumaatmaja@durham.ac.uk}
\affiliation{Department of Physics, Durham University, South Road, Durham DH1 3LE, UK}
 
\date{\today}

\begin{abstract}
A thermodynamically consistent free energy model for fluid flows comprising of one gas and two liquid components is presented and implemented using the entropic lattice Boltzmann scheme. The model allows high-density ratio, up to of order $O(10^3)$, between the liquid and gas phases; and a broad range of surface tension ratios, covering partial wetting states where Neumann triangles are formed, and full wetting states where complete encapsulation of one of fluid components is observed. We further demonstrate that we can capture the bouncing, adhesive and insertive regimes for the binary collisions between immiscible droplets suspended in air. Our approach opens up a vast range of multiphase flow applications involving one gas and several liquid components.

\end{abstract}

\maketitle

Multiphase flows comprising of one gas and 
several liquid components are of considerable scientific interest 
due to their broad range of applications. The collision between oil 
and water droplets is a key ignition step in combustion engines, 
where the collision parameters can be varied to control  the effective 
burning rate \cite{wang2004burning}. The presence of an immiscible 
crude oil layer on the sea surface alters the processes occurring during 
raindrop impact, with consequences for marine aerosol creation and oil 
spill dispersal \cite{murphy2015splash}. In advanced oil recovery, 
considerable gain can be achieved by alternately displacing the oil 
by air and water in the so-called immiscible water-alternating-gas
displacement process \cite{holtz2016immiscible}. 
Infusing porous materials with lubricants results 
in composite surfaces, known as lubricant impregnated surfaces 
\cite{Lafuma2011,smith2013droplet,Wong2011,Semprebon2017}, 
with superior non-wetting and drag-reduction properties.

% Explain the need
Despite the wide-ranging applications, suitable quantitative models 
for studying these phenomena are surprisingly still lacking. 
Most simulations to date have focussed on either single-component 
liquid-gas systems with high density ratio 
\cite{chikatamarla2015entropic,Brown2014,inamuro2004lattice,lee2005stable,Fakhari2017} 
or multicomponent flows with equal (or similar) density ratio 
\cite{Gunstensen1991,ShanChen1993,Blowey1996,Boyer2006,semprebon2016ternary, liang2016lattice,Dong2017}. 
In contrast, our aim here is to demonstrate an accurate and 
flexible model that can predict complex interfacial dynamics of ternary 
systems with significant ratio between the liquid and gas densities, 
up to of order $O(10^3)$. This enables a new class of multiphase problems 
to be simulated, which are not previously possible. While we focus on one gas and two 
liquid components, the model can be extended to include more liquid components.

% Why lattice Boltzmann for this problem
Our approach is based on the lattice Boltzmann method (LBM) 
\cite{succi2001,kruger2016lattice}, which has been shown to 
deliver reliable results, with quantitative agreement against 
experiments and other simulation methods, 
%in several areas of multiphase and multicomponent flows, 
including on droplet dynamics \cite{Liu2015,Lee2010,Varagnolo2013}, 
liquid phase separation \cite{Wagner1998,Kendon1999} and flow through porous media \cite{Liu2016}. 
In LBM interfacial forces can be implemented without explicit 
tracking of the interfaces, making it an elegant choice for 
studying mesoscopic interface dynamics in complex geometries. 
%LBM is also highly suitable for parallel and GPU computing \cite{Obrecht2013,schonherr_multi-thread_2011}, thus advantageous for computational problems with demanding time and length scales.
 
% Our contribution
Our key contribution over existing LBM models is a 
novel free energy functional that combines optimal equation of 
state for liquid-gas systems with double-well potentials to 
introduce multiple liquid components. The former, combined with 
the use of entropic lattice Boltzmann scheme \cite{chikatamarla2015entropic}, allows us to 
introduce significant density ratios, compared to other ternary 
free-energy LBM models \cite{semprebon2016ternary, liang2016lattice}. 
The free energy formulation also ensures our model is thermodynamically consistent, 
%an important feature for multicomponent fluids, 
unlike alternative approaches \cite{ShanChen1993,bao2013lattice,Li2014}. 

The capabilities of our new model are demonstrated using several
static and dynamic problems. Firstly, we find excellent agreement 
between the numerical and analytical liquid-gas coexistence curves 
as a function of temperature, proving the thermodynamic  
consistency of the model. Secondly, we illustrate how the liquid-liquid 
and liquid-gas surface tensions can be flexibly tuned by simulating 
liquid lenses with varying Neumann angles. Finally, we simulate 
binary collisions between two immiscible droplets and show 
we capture many relevant features reported in experiments
\cite{Wang2004,ChenChen2006,Chen2007,Roisman2012,Pan2016}.

\begin{figure}
\includegraphics[scale=0.24]{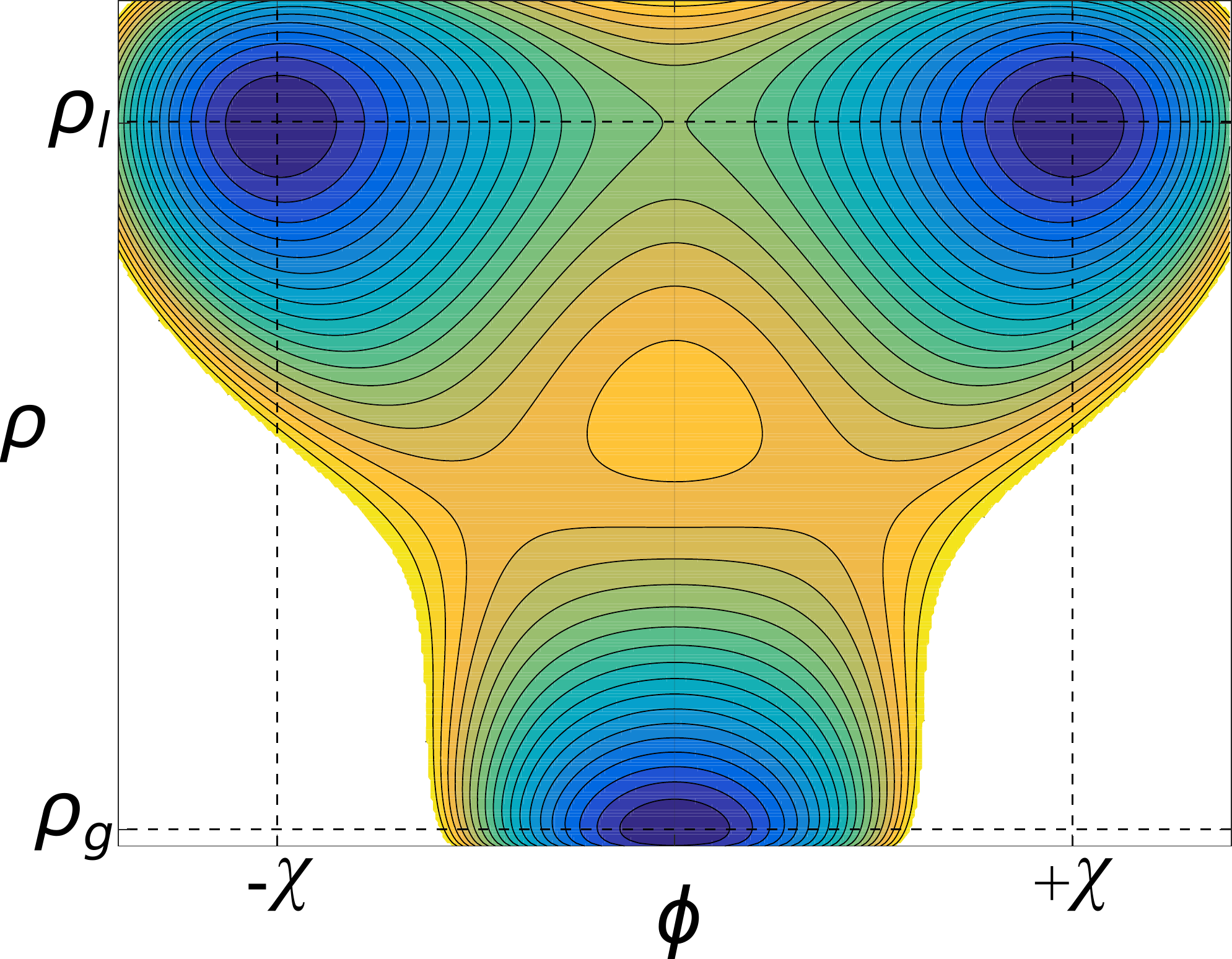}
\caption{Contour plot of the bulk free energy density $f_{\rm B}$ 
as a function of two order parameters, $\rho$ and $\phi$. 
Three distinct minima exist, corresponding to a gas component at 
$(\rho_g,0)$, and two liquid components at $(\rho_l,+\chi)$ and $(\rho_l,-\chi)$.}
\label{fig:fBulk}
\end{figure}

We introduce a free energy functional that consists of two parts, the bulk and interfacial contributions:
\begin{eqnarray}
F &=& \int \left[ f_{\rm B} + f_{\rm I} \right ] dV,  \label{equ:freeenergy}\\
f_{\rm B} &=& \frac{\lambda_1}{2} (\Psi_{\rm eos}(\rho) - \Psi_0) +  \label{equ:BulkEnergy} \\
&& \frac{\lambda_2}{2} C_{l1}^2(1-C_{l1})^2 +\frac{\lambda_3}{2} C_{l2}^2(1-C_{l2})^2, 
 \nonumber \\
f_{\rm I} &=& \frac{\kappa_1}{2} (\bm{\nabla} \rho)^2 
 + \frac{\kappa_2}{2}(\bm{\nabla} C_{l1})^2 
 +\frac{\kappa_3}{2} (\bm{\nabla} C_{l2})^2.  \label{equ:InterfaceEnergy}
\end{eqnarray}
The bulk free energy density $f_{\rm B}$ is designed to allow three distinct 
minima, corresponding to one gas and two liquid components, as illustrated in Fig. \ref{fig:fBulk}. 
$\Psi_{eos}(\rho)$ can be derived from integrating 
the liquid-gas equation of state (EOS), $p_{eos}=\rho(d \Psi_{eos} / d \rho)-\Psi_{eos}$, 
with coexisting liquid-gas densities at $\rho_l$ and $\rho_g$. 
For concreteness, here we use Carnahan-Starling EOS, but our approach
is flexible, and in SI \cite{supplementary} we describe the implementation of
Peng-Robinson and  van der Walls EOS. For Carnahan-Starling \cite{yuan2006equations}:
\begin{equation}
\label{equ:PsiEOS}
\Psi_{eos}=\rho\left(C-a\rho-\frac{8RT(-6+b\rho)}{(-4+b\rho)^2} +RT\log(\rho)\right).
\end{equation}
The constants $C$ and $\Psi_0$ are chosen such that 
$\Psi_{eos}(\rho_g)= \Psi_{eos}(\rho_l)=\Psi_0$, ensuring
common tangent construction is met between all 
coexisting fluid phases. We use $a=0.037$, $b=0.2$ and $R=1$. The critical 
temperature is $T_c=0.3373\frac{a}{bR}$, and the temperature $T$ governs 
the liquid-gas density ratio.

%{\color{red}Need non-LBM references.} 

The second and third terms in Eq. \eqref{equ:BulkEnergy} have the form of double well potentials with $C_{l1}$ and $C_{l2}$ the relative concentrations of the two liquid components. 
Established works on critical phenomena show that such form is universal to describe the physics of continuous phase transitions close to the critical point \cite{Cardy1996}, including for fluid mixtures. Away from the critical point, additional terms may be needed. However, so long as the details of the equation of state of the fluid mixtures is not important for the problem at hand, a large body of work in diffuse interface models for binary fluids has shown a double well potential is sufficient to capture interfacial dynamics with excellent agreement against both theory and experimental results, such as for droplet dynamics \cite{Liu2015,Lee2010} and coarsening in phase separation \cite{Wagner1998,Kendon1999}. This is the case for the examples studied here, and the double well potentials are therefore chosen as the simplest model possible.

Each double well term has two minima at $C_{l\#}=0$ (component absent) and 
$C_{l\#}=1$ (present). We also define the relative concentration 
of the gas phase as $C_g= (\rho-\rho_l)/(\rho_g-\rho_l)$, which is $0$ 
for $\rho = \rho_l$ and $1$ for $\rho = \rho_g$. Given the 
constraint $C_g + C_{l1} + C_{l2} = 1$, there are two independent 
order parameters: the density $\rho$ and 
the phase field $\phi$. The relative concentrations are related to the 
density and phase field via
$C_{l1} = \frac{1}{2} \left[1+\phi/\chi- (\rho-\rho_l)/(\rho_g-\rho_l)\right]$ 
and $C_{l2} = \frac{1}{2} \left[1-\phi/\chi- (\rho-\rho_l)/(\rho_g-\rho_l)\right]$,
with $\chi$ a constant scaling parameter for $\phi$. 
Our free energy functional has three minima at 
$(\rho_g,0)$, $(\rho_l,+\chi)$ and $(\rho_l,-\chi)$. 

For the interfacial free energy density, $f_{\rm I}$, all three terms in Eq. \eqref{equ:InterfaceEnergy} 
are necessary because there are three independent surface tensions in ternary systems.  
Upon expanding $C_{l\#}$ in terms of $\rho$ and $\phi$, we can rewrite $f_{\rm I}$ as
\begin{eqnarray}
 f_{\rm I} &=& \left[ \frac{\kappa_1}{2} +  \frac{\kappa_2+\kappa_3}{8(\rho_g-\rho_l)^2} \right] (\bm{\nabla} \rho)^2 + \frac{\kappa_2 +\kappa_3}{8\chi^2}(\bm{\nabla} \phi)^2 \nonumber \\
&& +\frac{\kappa_3 - \kappa_2}{4\chi(\rho_g-\rho_l)} (\bm{\nabla} \rho \cdot \bm{\nabla} \phi).
\end{eqnarray}
We vary the $\lambda$ parameters in Eq. \eqref{equ:BulkEnergy}
and $\kappa$ parameters in Eq. \eqref{equ:InterfaceEnergy} to tune the surface tensions
and interfacial widths of the three fluid interfaces.

The continuum equations of motion for the fluid are the 
continuity, Navier-Stokes, and Cahn-Hilliard equations:
\begin{eqnarray}
& \partial_t \rho + \bm{\nabla} \cdot \left( \rho \bm{v} \right) = 0, \label{eq:Continuity}  \\
& \partial_t (\rho \bm{v}) + \bm{\nabla} \cdot \left( \rho \bm{v} 
\otimes \bm{v} \right) = - \bm{\nabla} \cdot \bm{P} + \bm{\nabla} \cdot \left[ \eta ( \bm{\nabla v} +  \bm{\nabla v^T} ) \right], \,\,\,\,\,\,\, \label{eq:NS}  \\
& \partial_t  \phi + \bm{\nabla} \cdot (\phi \bm{v}) = M \nabla^2 \mu_\phi. \label{eq:Cahn-Hilliard} 
\end{eqnarray}
$\bm{v}$ is the fluid velocity and $\eta$ is the dynamic viscosity that depends on the local density and phase field. For simplicity, we employ constant mobility parameter $M$, though in general it can depend on the local density and phase field \footnote{
There has been several works dedicated to understanding the choice of the mobility parameter $M$, including on coarsening dynamics during phase separation \cite{Dai2016,Zhu1999}, droplet coalescence and breakup \cite{Jacqmin1999,Dupuy2010}, and contact line motion of a fluid interface at a solid boundary \cite{Kusumaatmaja2016,Yue2010}.}.
The thermodynamics of ternary fluids, described by the free energy functional $F$ in Eq. \eqref{equ:freeenergy},
enter the equations of motion through the chemical 
potentials, $\mu_{\rho}=\delta F/\delta \rho |_{T,\phi}$ and 
$\mu_{\phi}=\delta F/\delta \phi |_{T,\rho}$, and the pressure 
tensor, $\bm{\nabla} \cdot \bm{P} = \rho \bm{\nabla} \mu_{\rho} + \phi \bm{\nabla} \mu_{\phi}$. 
To solve the equations of motion, we introduce two sets of distribution functions
in our LBM scheme, evolving the density and phase field. For the former, we employ 
the entropic lattice Boltzmann method, augmented with an 
exact-difference forcing term \cite{chikatamarla2015entropic}. 
For the latter we use a standard BGK scheme \cite{kruger2016lattice}.
We provide the details of our LBM implementation in SI \cite{supplementary},
including the expressions for the chemical potentials and pressure tensor.

%%%%%%%%%%%%%%%%%%%
\begin{figure*}[t]
\includegraphics[width=0.92\linewidth]{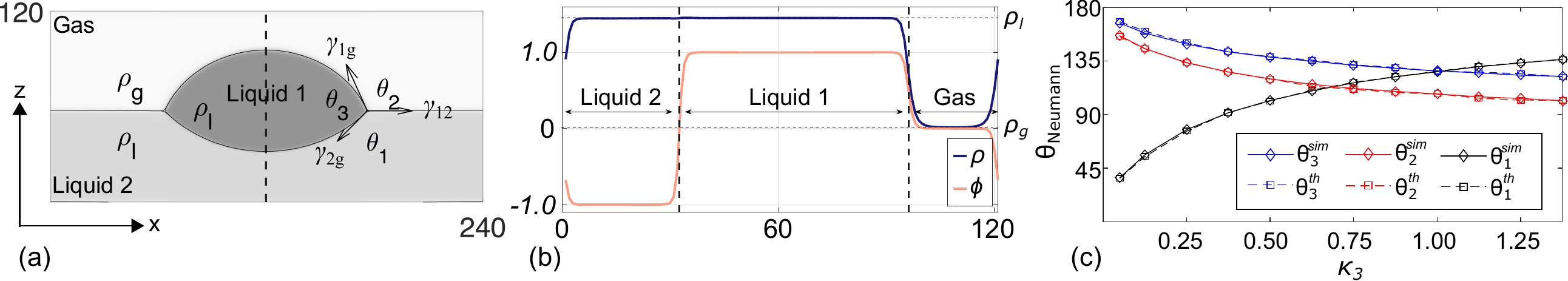}
\caption{ (a) Simulation of a liquid 1 droplet suspended at the interface 
between liquid 2 and the gas phase. The Neumann angles $\theta_1, \theta_2, \theta_3$ 
are a consequence of mechanical equilibrium between the interfacial 
tensions. (b) The variation of the density $\rho$ 
and phase-field $\phi$ across the vertical dashed line shown in 
(a). (c) Variation of the Neumann angles by varying $\kappa_3$. We fix 
$\lambda_3=3.125\times\kappa_3$, $\lambda_1 = 0.6$, $\kappa_1 = 10^{-3}$, $\kappa_2 = 1.0$, $\lambda_2 = 3.125\times\kappa_2$, $T_r = T/T_c = 0.69$ and $\chi=5.0$.
}
\label{fig:LL}
\end{figure*}
%%%%%%%%%%%%%%%%%%%

To demonstrate the accuracy and broad range of surface tension ratios allowed in our model, 
we simulate a liquid lens, where a droplet of liquid 1 is suspended at 
the interface between liquid 2 and the gas phase, as shown in Fig. \ref{fig:LL}(a). 
We show the profiles of 
the density $\rho$ and phase field $\phi$ across the liquid lens configuration 
in Fig. \ref{fig:LL}(b). At the interface between liquids 1 and 2, $\rho$
remains constant at $\rho_l$, while $\phi$ transforms 
smoothly between $-\chi$ and $\chi$. Both $\rho$ and $\phi$ vary
at the interface between any of the liquids and the gas.
%These observations conform with the prescriptions of our free energy functional.

At equilibrium, force balance between the surface tensions at the 
three phase contact line leads to a distinct set of angles 
known as the Neumann angles. Mathematically, 
$\gamma_{12}/\sin(\theta_3)=\gamma_{2g}/\sin(\theta_2)=\gamma_{1g}/\sin(\theta_1)$. 
To test this relation we vary the value of $\kappa_3$, while keeping 
$\lambda_3=3.125 \times \kappa_3$ and other simulation parameters 
(see caption of Fig. \ref{fig:LL}) constant. 
For all simulations shown here, we also set the kinematic viscosity
$\nu = \eta/\rho = 0.167$, and mobility parameter $M = 0.5$.
Fig. \ref{fig:LL}(c) shows the Neumann angles calculated in two different ways. Firstly, 
we measure the Neumann angles geometrically (diamond symbol) 
from our liquid lens simulations. Secondly, we use Laplace pressure tests to independently 
measure surface tensions for all permutations of the interfaces (see SI \cite{supplementary}),
and subsequently compute the expected Neumann angles (square symbol). 
The agreement is excellent, with typical 
deviations of $< 3^\circ$. Similar agreement is observed upon varying 
other parameters.

In addition to partial wetting states, where the Neumann triangle is 
formed, our model allows simulations of full wetting states. 
To demonstrate this, in SI \cite{supplementary}, we present simulation results of two 
droplets where $\gamma_{1g} + \gamma_{12} < \gamma_{2g}$. 
The droplets are initialised such that they are just touching each other. 
As dictated by thermodynamics, the simulation shows that the 
liquid 2 droplet becomes fully encapsulated by the liquid 1 droplet. 

\begin{figure}[b]
\includegraphics[scale=0.23]{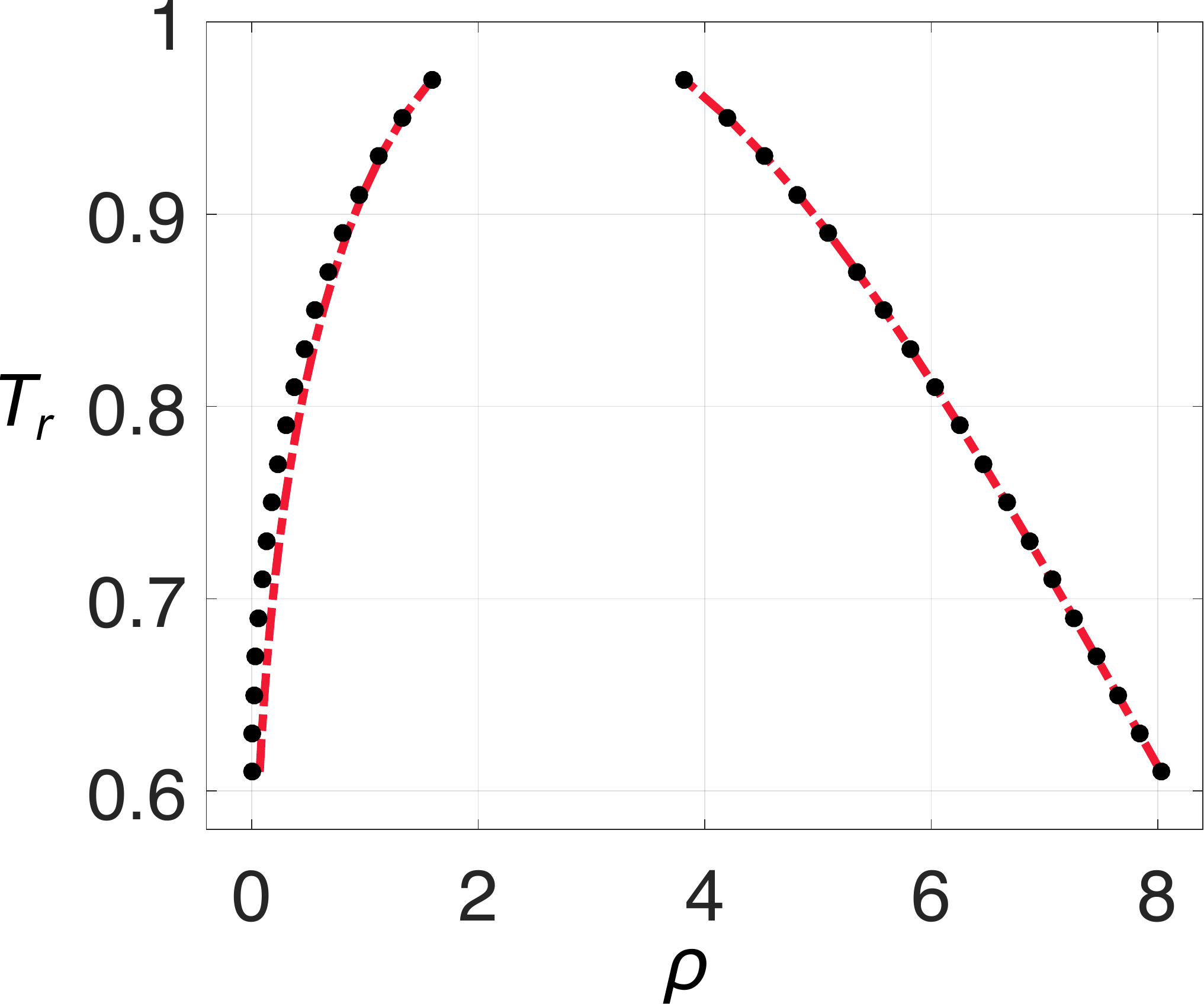} 
\caption{The coexistence curve in the numerically stable regime for 
Carnahan-Starling EOS as a function of the reduced temperature $T_r=T/T_c$.
}
\label{fig:Coex}
\end{figure}

A wide range of density ratios can be simulated 
by tuning the temperature $T$ in the equation of state, Eq. \eqref{equ:PsiEOS}. 
Fig. \ref{fig:Coex} shows the coexistence curve for the Carnahan-Starling EOS. 
The left (right) branches correspond to the gas (liquid) densities.
% while the right branch the liquid density. 
Good agreement is obtained between the analytical solution from 
Maxwell construction (line) and the numerical results (dots). The lowest 
temperature we can robustly simulate is 
$T=0.61T_c$, corresponding to a numerical density ratio of $O(10^3)$.
In SI \cite{supplementary} we also show that high density ratios can be achieved with
Peng-Robinson and van der Walls EOS.

We now present simulation results of collisions between two immiscible droplets.
In comparison to the more commonly studied problem of collisions between miscible droplets 
of the same materials (e.g. \cite{chikatamarla2015entropic,Brown2014}), 
the collision outcomes for immiscible droplets are much richer. 
Here we show three regimes observed in experiments: bouncing, adhesive and insertive collisions, 
and their transitions. To our best knowledge, this is the first time they have 
been simulated using LBM. We will focus on generic features of the 
drop collisions. Systematic studies, including parameter matching against 
experiments, will be presented elsewhere.

\begin{figure}[] 
\includegraphics[scale=0.65]{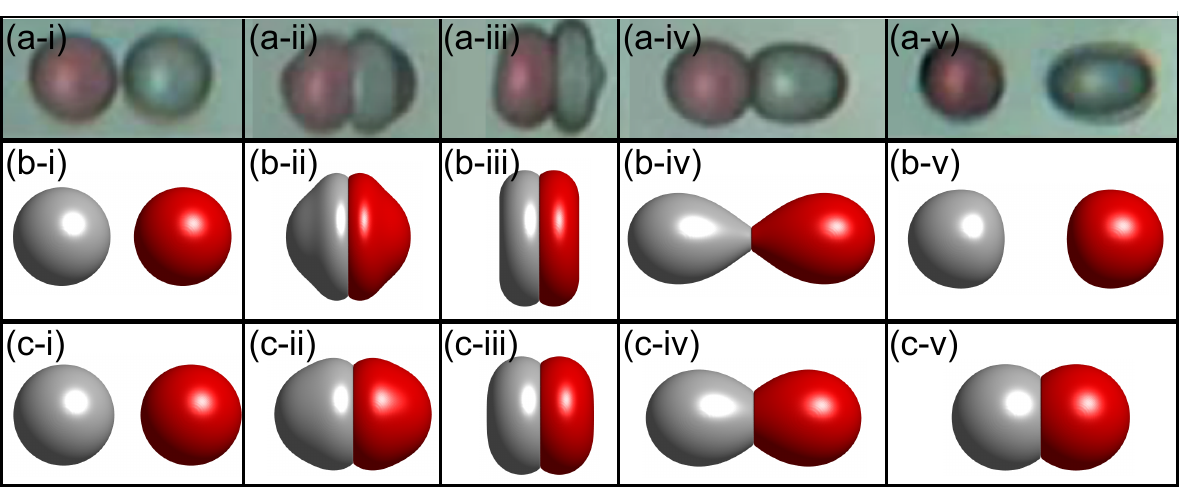} 
\caption{(a) An experimental example of bouncing collision between two 
immiscible droplets (water and diesel oil) \cite{ChenChen2006}. (b-c) Transition from bouncing 
(b) to adhesive (c) collisions can be induced by decreasing the 
droplets' velocities.}
\label{fig:Adhesive}
\end{figure}

We first consider bouncing collision. Fig. \ref{fig:Adhesive}(a) shows an 
experimental example where the two droplets are water and diesel oil \cite{ChenChen2006}. 
As the droplets collide (columns ii and iii), we observe compression in 
the drop shapes parallel to the collision direction and radial expansions 
perpendicular to the collision direction. This is followed by retraction 
in the radial direction (column iv), and if there is sufficient kinetic energy, 
the two droplets bounce off and become separated (column v). Our simulations 
show this sequence is ubiquitous for head-on bouncing collisions.
Fig. \ref{fig:Adhesive}(b) shows one such case at 
${\rm{We}}_1 = {\rm{We}}_2 = 20.8$ and ${\rm{Re}}_1 = {\rm{Re}}_2 = 72.0$, 
where ${\rm{We}}_i = \rho_i V_r^2 D_i/\gamma_{ig}$, 
${\rm{Re}}_i = \rho_i V_r D_i / \eta_i$, and $V_r$ is the relative droplet velocity. 
Here the two droplets have symmetric properties and we use  
$\gamma_{12}/\gamma_{2g} = 1.33$. We set $T = 0.65 T_c$ for the rest of the paper, corresponding to
a density ratio of $\rho_l/\rho_g \simeq 150$.
For the cases shown here, the results do not sensitively depend
on the density ratio beyond $\rho_l/\rho_g \sim 100$. This is illustrated explicitly in SI \cite{supplementary} by comparing the results in Fig. \ref{fig:Adhesive} to 
those obtained using $T_r = 0.61$ ($\rho_l/\rho_g \simeq 1000$).

By reducing the droplets' velocities, we observe a transition 
from bouncing to adhesive collision, shown in Fig. \ref{fig:Adhesive}(c) 
for ${\rm{We}}_1 = {\rm{We}}_2 = 5.6$ and ${\rm{Re}}_1 = {\rm{Re}}_2 = 36.0$. 
Qualitatively the initial collision dynamics is similar 
between rows (b) and (c). However, at column (iv) there is not enough kinetic 
energy for the droplets to detach. Subsequently 
the compound droplet oscillates until it relaxes to its equilibrium 
configuration, determined by the Neumann triangle. 
Animations of the drop collisions in Fig. \ref{fig:Adhesive}(b) and (c) are provided as SI \cite{supplementary}.
Adhesive collision between two immiscible droplets with similar liquid-gas surface tension
has been observed experimentally for diesel and ethanol droplets \cite{Chen2007}.

\begin{figure}[] 
\includegraphics[scale=0.65]{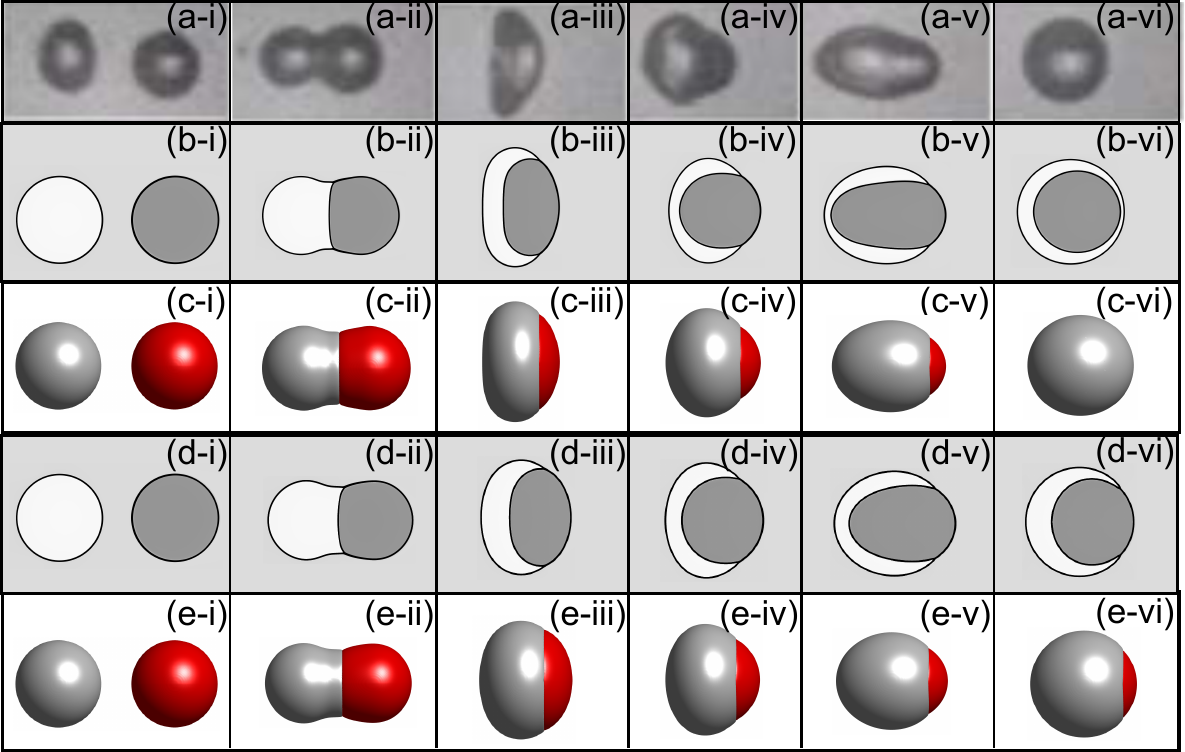} 
\caption{(a) An experimental example of insertive collision between two 
immiscible droplets (water and n-hexadecane) \cite{Wang2004}, where one droplet is fully 
encapsulated by the other. Transition from insertive (b-c) to adhesive 
(d-e) collisions can be induced by decreasing the droplets' velocities. 
Rows (b) and (d) are cross-sections of the drops in rows (c) and (e).}
\label{fig:Insertive}
\end{figure}

A powerful advantage of our model is that it covers a wide range of surface 
tension ratios. We can now consider the asymmetric case where the 
liquid-gas surface tension of droplet 2 is considerably larger than that for 
droplet 1, yet it does not correspond to the full wetting state. 
Fig. \ref{fig:Insertive}(b-e) shows the case where $\gamma_{12}/\gamma_{2g} = 0.54$ 
and $\gamma_{1g}/\gamma_{2g} = 0.49$, with normalised spreading parameter 
$S = 1 - (\gamma_{1g} + \gamma_{12})/ \gamma_{2g} = - 0.029$. In agreement 
with experimental observations \cite{Wang2004}, we observe a transition between 
adhesive and insertive collisions by varying the impact velocities. 

An experimental example of insertive collision is shown in Fig. \ref{fig:Insertive}(a) 
for water and n-hexadecane \cite{Wang2004}. For comparison, Fig. \ref{fig:Insertive}(b) and (c) show 
the typical dynamical sequence observed in our simulations, with
${\rm{We}}_1 = 16.4$, ${\rm{We}}_2 = 6.1$ and ${\rm{Re}}_1 = {\rm{Re}}_2 = 45.0$.
Upon collision, the composite droplet expands radially (column iii), followed by 
contraction in the radial direction (column iv) and elongation in the collision 
axis (column v). The oscillation between the prolate and oblate shapes can sustain 
several periods (videos in SI \cite{supplementary}), accompanied 
by the propagation of the three-phase contact line until the high surface tension 
droplet is fully encapsulated (column vi). 

The transition from insertive to adhesive collision can be induced by decreasing 
the droplets' velocities. In Fig. \ref{fig:Insertive}(d) and (e), we present the 
case where ${\rm{We}}_1 = 4.4$, ${\rm{We}}_2 = 2.2$ and ${\rm{Re}}_1 = {\rm{Re}}_2 = 27.0$.
Initially the contact line propagates to cover the high surface tension droplet 
as the composite droplet oscillates between the prolate and oblate shapes (videos in SI \cite{supplementary}). 
Since the kinetic energy is insufficient to drive full encapsulation, 
the contact line eventually recedes and the droplet relaxes to its equilibrium shape (column vi).
%, again following the Neumann triangle. 
In SI \cite{supplementary} we further show the critical velocity for 
the transition between insertive and adhesive collisions increases as 
the normalised spreading parameter becomes more negative.

%The critical We and Re for the transition between insertive and adhesive 
%collisions depend on the spreading parameter. For instance, lowering it to 
%$S = -0.04$, insertive collisions are only observed (see SI) for Re and We 
%larger than ${\rm{We}}_1 = 5.86$, ${\rm{We}}_2 = 2.19$, ${\rm{Re}}_1 = {\rm{Re}}_2 = 27.00$.

To conclude, we presented a strategy for modelling ternary multiphase multicomponent flows
 by combining a novel free energy formulation and the use of entropic LBM scheme. 
 Our approach allows significant density ratios, up to of 
 order $O(10^3)$, and a broad range of surface tension ratios, covering both 
 partial and full wetting states, to be simulated. These flexibilities 
 open up a number of applications, which are not previously possible.
 As an example, we demonstrated the bouncing, adhesive and insertive regimes for binary collisions between 
 immiscible droplets. Our method can meet the gap in systematic computational 
 work for such collision dynamics, to complement the rich body of existing 
 experimental studies \cite{Wang2004,ChenChen2006,Chen2007,Roisman2012,Pan2016}. 
Other applications are numerous,
including drop impact on immiscible liquid film \cite{murphy2015splash},
advanced oil recovery \cite{holtz2016immiscible},  and liquid impregnated surfaces
\cite{Lafuma2011,smith2013droplet,Wong2011,Semprebon2017}.
%, and self-propulsion of liquid bi-slug \cite{Bico2002}.}  
  
Here we have assumed the liquids to have the same density. This is justifiable 
in most water-oil-gas systems where the liquid-liquid density ratio is several 
orders of magnitude smaller than the liquid-gas density ratio. A 
useful future extension is to allow all density ratios to be varied independently. 
Our model can also be generalised to include more liquid components, by introducing additional double well 
potential and gradient terms in the bulk and interfacial free energy densities respectively.
Another key avenue for future work is the interactions between ternary 
flows and complex solid surfaces. Our model is compatible with 
various approaches to introduce wetting boundary conditions \cite{Ding2007,Huang2007,Huang2015}.
%(e.g. geometric, force, surface energy) 
 %, and it will be interesting to investigate which boundary condition is superior in this context. 

%(see discussion in SI)

\begin{acknowledgments}
We acknowledge funding from Procter \& Gamble (HK), EPSRC (HK; grant EP/P007139/1) and SNF (IK; grant 200021\_72640).
\end{acknowledgments}

%#####################################
%#####################################
\bibliographystyle{apsrev4-1}
\bibliography{Ref}

\end{document}